\documentstyle[amsfonts,12pt]{article}
\def\one{1\hskip-.37em 1}

\def\half{\textstyle{\frac{1}{2}}}

\def\H{{\cal H}}

\def\ra{\rightarrow}
\def\tint{{\textstyle\int}}

\def\s{\hskip.08em}

\def\b{\begin{eqnarray*}}     %takes no eqn numbers
\def\e{\end{eqnarray*}}       %takes no eqn numbers
\def\bn{\begin{eqnarray}}     %takes eqn numbers 
\def\en{\end{eqnarray}}       %takes eqn numbers
\def\<{\langle}
\def\>{\rangle}

\def\{{\lbrace}
\def\}{\rbrace}
\begin{document}
%\footnote{Electronic mail: klauder@phys.ufl.edu}
\title{The Current State of Coherent States\footnote{Contribution to 
the 7th ICSSUR Conference, June 2001.}}
\author{John R. Klauder\\
%\footnote{Electronic mail: klauder@phys.ufl.edu}\\
Departments of Physics and Mathematics\\
University of Florida\\
Gainesville, FL  32611}
%Email: klauder@phys.ufl.edu}
\date{}     %   Use   %\date{} to see the dates
\maketitle
\begin{abstract}
The original canonical coherent states could be defined in several ways. 
As applications for other sets of coherent states arose, the rules of 
definition were correspondingly changed. Among such rule changes were 
a change of group and relaxation of the analytic nature of the labels. 
Recent developments have done away with the group connections altogether 
and thereby allowed sets of coherent states to be defined that are 
temporally stable for a wide variety of dynamical systems including 
the hydrogen atom. This article outlines some of the current trends 
in the definitions and properties of present-day coherent states.
\end{abstract}
%\vfill\eject
\section*{Introduction}
The modern reincarnation of what are now often called canonical coherent 
states began in 1960 \cite{kla2} (with a mathematical-physics application 
to define coherent state path integrals), in 1961 \cite{barg} (with a 
thorough mathematical study), and in 1963 \cite{gla} (with a physics 
application central to the new theory of quantum optics). Over the years, 
generalizations of the original family of canonical coherent states have 
been introduced based largely on mathematical or possibly 
mathematical-physics grounds. These generalizations have frequently 
involved one or another of the mathematical properties of the canonical 
coherent states and its elevation to the central concept in defining 
new sets of coherent states. As examples, we cite group-defined coherent 
states \cite{kla19,per,gil}, annihilation-operator-eigenstate defined 
coherent states \cite{bar}, and minimum-uncertainty-state defined 
coherent states \cite{nie}. Such generalizations typically lead to new 
sets of coherent states alright, but (apart perhaps from the 
group-defined coherent states) such rules for generating new sets of 
coherent states have always seemed to the present author to be overly 
mathematical and rather divorced from any specific physics. After all, 
what is the physics involved in choosing annihilation-operator 
eigenstates or in choosing minimum uncertainty states? What would be so 
wrong in choosing states for which the minimum uncertainty product was 
exceeded by a factor of three, for example?

These views have in recent years prompted the author to seek other 
generalizations of the canonical coherent states often with specific 
physical criteria chosen as the key factor involved in defining and 
obtaining such generalizations. Although other prescriptions exist, we 
shall, in the interests of brevity and consistency, pursue just one path 
among many in our discussion of new sets of coherent states.

A few introductory remarks are useful: For convenience, we denote each of 
the coherent states by $|l\>\in{\frak{H}}$,
$|l\>\ne0$, where $l=(l^1,l^2,\ldots,l^L)\in{\cal L}$, $l^j\in{\mathbb R}$,
denotes an $L$-dimensional (real) label lying in a label space $\cal L$ 
which locally is
topologically equivalent to ${\mathbb R}^L$. This latter property means 
that we can identify
continuous functions on $\cal L$. It is often useful to regard $l$ as 
a {\it classical} variable in a classical (phase) space $\cal L$. Although 
we shall not generally do so, it is often useful to group 
some or all of the real parameters by pairs and to form complex parameters. 
Throughout, we choose units so that $\hbar=1$.

With these remarks as background, we start with what we regard as the basic 
minimum properties for any set of states to be called a set of coherent 
states:
\vskip.4cm

{\bf 1.~ Continuity of Labeling:} The map from the label space $\cal L$ 
into the Hilbert space $\frak H$ is strongly continuous. 

{\bf Comment:} Specifically, this condition requires that the expression 
$\| |l'\>-|l\>\|\ra0$ whenever $\l'\ra l$ in $\cal L$. This condition is 
equivalent to the joint continuity of the coherent state overlap function, 
$\<l''|l'\>$, in its two arguments. \vskip.4cm

{\bf 2. Resolution of Unity:} A positive measure $\mu(l)$ on $\cal L$ exists 
such that the unit operator $\one$ admits the representation
   $$ \one = \int_{\cal L}|l\>\<l|\,d\mu(l)\;,  $$
where $|l\>\<l|$ denotes the rank-one operator that takes an arbitrary vector
$|\psi\>$ into a multiple (namely $\<l|\psi\>$) of the vector $|l\>$.

{\bf Comment:} If $|l\>=0$ for some $l$, these vectors would make no 
contribution to the resolution of unity, and so we have already assumed 
that $|l\>\ne0$,
i.e., $\<l|l\>>0$. If $d\mu(l)=0$ for a set of nonzero measure, then 
these vectors would also not contribute to the resolution of unity. Hence, 
there is no loss of generality to require that $\mu(l)$ is a strictly 
positive measure (up to sets of measure zero). In addition, it is often 
useful to assume that $\mu(l)$ is scaled (or rescaled, if necessary) so 
that $\<l|l\>=1$ for all $l\in{\cal L}$. If $\<l|l\>=1$, then it follows 
that $|l\>\<l|$ is a one-dimensional projection operator. (Ideally, 
$\mu(l)$ should be a countably additive measure, but a finitely additive 
measure is generally sufficient, which is a distinction for positive 
measures that may arise when an infinite number of degrees of freedom 
are involved.)

{\bf Remark:} The two postulates about coherent states above were proposed 
in substantially this form nearly forty years ago \cite{kla18}, even before 
such states were called ``coherent states''. With very few exceptions, all 
states that have been so named have fulfilled these two postulates and for 
purposes of the present article we shall require that these two postulates 
hold. [For a recent study of a case where a resolution of unity 
(Postulate 2) fails to hold, see \cite{kla222}.]

The generality of the first two postulates, and their mathematical 
specificity as well, has been done deliberately so that a vast catalog 
of sets of coherent states implicitly exists; ideally, it is the analysis 
of a specific physical problem which, whenever possible, puts on additional 
physical restrictions that singles out a subset of coherent-state 
sets---or even a single coherent-state set---tied to the specified 
physical problem.

A useful analogy to the present point of view lies in the mathematical 
concept of a set of {\it orthonormal functions}. Initially, one can define 
the properties that make a set of functions an acceptable set of orthonormal 
functions, i.e., completeness, orthogonality, and normalization. Finally, 
one can introduce criteria to select some sets or even one set of 
orthonormal functions relevant to some specific physical problem.

{\bf Remark:} The philosophy of defining coherent states expressed here is, 
of course, just one of many possible choices. Others are free to choose 
alternative definitions, although it naturally diminishes the utility of 
the phrase when it is used too widely. The ultimate value of a definite 
rule of definition stems from its {\it usefulness in applications}$\s$; 
and applications generally arise for specific and concrete systems.

We now turn our attention to picking out suitable sets or even a single set 
of coherent states by adopting certain physical criteria rather than 
imposing selected mathematical requirements as discussed in the previous 
section. 

\section*{Temporal Stability}
For our first additional property we shall study time evolution as 
dictated by a specific Hamiltonian operator $\H$. The evolution of any 
coherent state $|l\>$ may always be captured by the relation
  $$  e^{-i\H t}\,|l\>\equiv |l,t\>\;,  $$
a definition that imposes no restriction whatsoever. However, we can ask 
for much more. Let us first restrict attention to normalized coherent 
states, $\<l|l\>=1$, for all $l\in{\cal L}$. Then we may ask that the 
following condition holds:
\vskip.4cm

{\bf 3. Temporal Stability:} The time evolution of any coherent state 
always remains a coherent state. In symbols,
  $$ e^{-i\H t}\,|l\>=|l(t)\>  $$
for all $\l\in{\cal L}$ and all $t\in{\mathbb R}$, 
where $l(0)=l$.\footnote{Although temporal stability 
refers to the {\it quantum} evolution of the coherent states, 
there is nonetheless an induced {\it classical} dynamics inherent 
in this concept that realizes the label-space map $l\ra l(t)$ for 
each $l\in{\cal L}$. We shall touch on this classical dynamics below.}

{\bf Comment:} In order to avoid any time-dependent scale factors, it 
has been useful to first assume that all coherent states are normalized. 
While the set of coherent states satisfies temporal stability, the same 
cannot be said for the temporal evolution of a general state 
$e^{-i\H t}\,|\psi\>\equiv|\psi,t\>$. Nevertheless, in a coherent 
state representation that enjoys temporal stability, {\it dynamics 
becomes kinematics}. In other words,
  \b  \<l|\psi,t\>\equiv\<l|\s e^{-i\H t}\s|\psi\>\equiv\<l(-t)|\psi\>\;, \e
namely, the dynamical evolution of an arbitrary state 
$\psi(l)\equiv\<l|\psi\>$ in the coherent state representation 
simply amounts to a ``reshuffling of the labels'', 
$\psi(l,t)\equiv\psi(l(-t))$.   
 
Let us see how we can explicitly implement temporal stability. 
For convenience, we restrict attention to Hamiltonians with a discrete, 
nondegenerate spectrum and energy levels of the form 
$0=E_0<E_1<E_2<\cdots\,$. It follows that $\lim_{n\ra\infty}E_n=E^*$, 
and cases where $E^*=\infty$ and $E^*<\infty$ are both of interest. 
We set $e_n\equiv E_n/\omega$, for some convenient choice of $\omega$, 
to generate a sequence of dimensionless energy levels. If $E^*<\infty$, 
we can, without loss of generality, choose $\omega=E^*$ so that 
$\lim_{n\ra\infty}e_n=1$. Furthermore, we let $|n\>$, 
$n=0,1,2,\ldots\,$, be energy eigenvalues for $\H$, such that
 \b  \H\s|n\>=E_n\s|n\>=\omega\s e_n\s|n\>\;.  \e
We then define (see \cite{kla191}) coherent states asociated with 
this system by the expression
\b |J,\gamma\>\equiv N(J)^{-1/2}\sum_{n=0}^\infty\frac{J^{n/2}\s 
e^{-ie_n\gamma}}{\sqrt{\rho_n}}\,|n\>\;,  \e
where $0\le J<J^*\le\infty$ and $-\infty<\gamma<\infty$, expressed 
with the aid of a set of positive weight factors $\{\rho_n\}$, with 
$\rho_0\equiv 1$ for convenience. Here normalization is achieved by setting
 \b N(J)=\sum_{n=0}^\infty\frac{J^n}{\rho_n}\;,  \e
and where $J^*\equiv \liminf_{n\ra\infty}[\rho_n]^{1/n}$ denotes the 
radius of convergence of this series. We note first that
\b  &&e^{-i\H t}\s|J,\gamma\>\equiv N(J)^{-1/2}\sum_{n=0}^\infty
\frac{J^{n/2}\s e^{-ie_n\gamma-i\omega e_nt}}{\sqrt{\rho_n}}\,|n\> \\
  &&\hskip2cm=|J,\gamma+\omega t\>  \e
whatever the choice of the weight factors $\{\rho_n\}$. Thus by a careful 
choice of the phase factor we have ensured temporal stability.

Let us next discuss the freedom in the choice of the factors $\{\rho_n\}$ so 
that the coherent states fulfill Property 2 dealing with the resolution of 
unity. To that end, we assume there exists a nonnegative weight function 
$\rho(u)$, $\rho(u)\ge0$, $0\le u<U\le\infty$, with the property that
  \b  \rho_n\equiv \int_0^Uu^n\s\rho(u)\,du\;;\hskip1.5cm \rho_0=1\;. \e
Next we observe that
 \b &&\int |J,\gamma\>\<J,\gamma|\,d\nu(\gamma)\equiv 
\lim_{\Gamma\ra\infty}(2\Gamma)^{-1}\s\int_{-\Gamma}^\Gamma 
|J,\gamma\>\<J,\gamma|\,d\gamma \\
  &&\hskip3.5cm=N(J)^{-1}\sum_{n=0}^\infty\frac{J^n}{\rho_n}\,|n\>\<n|\;.  \e
Finally, if we introduce $k(J)\equiv N(J)\s\rho(J)$ and $U\equiv J^*$, 
then we find that
 \b&&\hskip-.8cm\int |J,\gamma\>\<J,\gamma|\,d\mu(J,\gamma)\equiv
\int_0^U k(J)\,dJ\int d\nu(\gamma)\,|J,\gamma\>\<J,\gamma| \\
  &&\hskip1cm=\sum_{n=0}^\infty\frac{|n\>\<n|}{\rho_n}\,\int_0^U 
J^n\s\rho(J)\,dJ \\
&&\hskip1cm=\sum_{n=0}^\infty|n\>\<n|\equiv \one \;.  \e

As a result of this analysis, we learn that there are a vast number of 
coherent state sets, all of which fulfill temporal stability for a 
single Hamiltonian, and which are distinguished from each other by the 
presence of different weight factor sets $\{\rho_n\}$. 

We now seek an additional physical criterion that picks out a {\it single} 
set of weights $\{\rho_n\}$ for a given Hamiltonian, thereby reducing 
the vast family of coherent states down to a single set.

\section*{The Action Identity}
Let us return to the appropriate label map $l\ra l(t)$ for the set of 
coherent states under discussion. Specifically, the appropriate map in 
the present case is clearly given by $(J,\gamma)\ra(J,\gamma+\omega t)$. 
This temporal evolution is the most general solution of the two equations 
of motion
  \b  {\dot\gamma}=\omega\;, \hskip1.5cm {\dot J}=0\;,  \e
which in turn arise, for example, from the ``classical action functional''
 \b  I=\tint[J\s{\dot\gamma}-\omega\s J]\,dt  \e 
as the relevant Euler-Lagrange equations. In point of fact, other action 
functionals would work just as well, say, for instance, 
 \b  I'=\tint[J^3\s{\dot\gamma}-\omega\s J^3]\,dt  \;.  \e
However, there is an additional sense in which $I$ is preferred since in 
that case $J$ and $\gamma$ can be said to be {\it classical canonical 
coordinates}; this
interpretation is not supported by using $I'$ (or any other such form). 
Let us accept the physical notion that $J$ and $\gamma$ should represent 
classical canonical coordinates and thus $I$ corresponds to the 
appropriate classical action.

It is a longstanding proposal \cite{kla19,kla37} that there is just 
{\it one} action principle in physics, and that in particular, 
{\it the classical action principle is just the quantum action principle 
applied to a restricted set of Hilbert space vectors}. We can illustrate 
this proposal as follows: Let
\b I_Q=\tint[i\<\psi(t)|\s(d/dt)\s|\psi(t)\>-\<\psi(t)|\s\H\s
|\psi(t)\>]\,dt \e
denote the usual quantum action functional. Extremizing this functional 
over all bra vectors $\<\psi(t)|$ leads to Schr\"odinger's equation
 \b i(d/dt)\s|\psi(t)\>=\H\s|\psi(t)\>\;. \e
Let us ask the question, however, what is the result if we extremize the 
quantum action functional over a {\it limited set} of vectors such as 
those in a set of coherent states. For example, consider states of the form
  \b  |p,q\>\equiv e^{-iqP}\s e^{ipQ}\s|0\>\;,  \e
where $[Q,P]=i\one$ and $|0\>$, say, is a unit vector which 
satisfies $(Q+iP)\s|0\>=0$. It is then straightforward to show that
 \b &&I_Q=\tint[i\<p(t),q(t)|\s(d/dt)\s|p(t),q(t)\>-\<p(t),q(t)|\s\H\s
|p(t),q(t)\>]\,dt\\
&&\hskip.59cm=\tint[p(t)\s{\dot q}(t)-H(p(t),q(t))]\,dt\;, \e
where $H(p,q)\equiv\<p,q|\s\H\s|p,q\>$ is a classical Hamiltonian 
symbol asociated with the quantum Hamiltonian $\H$. Clearly, extremal 
variation of $I_Q$ within the limited set of coherent states, i.e., for 
general functions $p(t)$ and $q(t)$, leads to traditional classical 
equations of motion for the canonical variables $p$ and $q$. In this 
interpretation, classical dynamics is what remains of quantum dynamics 
when the latter is subject to a sufficiently large class of constraints 
that restrict possible variations. Stated otherwise, classical dynamics 
is quantum dynamics restricted to the only quantum degrees of freedom that 
may possibly be varied at a macroscopic level, namely, the mean position 
and the mean momentum (or velocity).

The foregoing discussion can be applied to the problem at hand as follows: 
If we seriously wish to identify the variables $J,\gamma$ of the coherent 
states $
|J,\gamma\>$ as canonical coordinates, then it is necessary that
  \b &&I=\tint[J\s{\dot\gamma}-\omega\s J]\,dt \\
  &&\hskip.35cm=\tint[i\<J,\gamma|\s(d/dt)\s|J,\gamma\>-\<J,\gamma|
\s\H\s|J,\gamma\>]\,dt\;.  \e
Consequently, we are led to the next, and last, postulate, namely\vskip.4cm

{\bf 4. Action Identity:} To ensure that the variables $J$ and $\gamma$ 
correspond to physical canonical coordinates, we require that
  \b \<J,\gamma|\s\H\s|J,\gamma\>=\omega\s J\;.  \e

{\bf Comment:} As easily seen, this last condition is equivalent to 
requiring that 
$i\<J,\gamma|\s d\s|J,\gamma\>=J\,d\gamma$. The action identity is a strong 
requirement, and we next show that it will uniquely specify the weight 
factors $\{\rho_n\}$ for a given Hamiltonian $\H$. 

The action identity asserts, for all $J$, $0\le J< J^*$, that
\b  \sum_{n=o}^\infty\frac{e_n\s J^n}{\rho_n}=J\s\sum_{n=0}^\infty
\frac{J^n}{\rho_n}\;.  \e
Equating like powers of $J$, we are led to the condition $e_n/\rho_n=
1/\rho_{n-1}$, or $\rho_n=e_n\s\rho_{n-1}$. Choosing $\rho_0=1$ 
(as already noted), we find that
  \b  \rho_n\equiv e_n\cdot e_{n-1}\cdots e_1 =\Pi_{l=1}^n\s e_l\;. \e
The final result is, therefore, the set of coherent states introduced by 
Gazeau and Klauder \cite{kla207}. 

It is instructive to apply the final coherent-state prescription to a 
familiar example, namely, to the harmonic oscillator. In that case, 
$E_n=\omega\s n$, or $e_n=n$, and so $\rho_n=n!$. If we let 
$|z|\equiv J^{1/2}$ and set $z\equiv|z|\s e^{-i\gamma}$, then we find that
 \b |J,\gamma\>\equiv|z\>=e^{-\half|z|^2}\s\sum_{n=0}^\infty
\frac{z^n}{\sqrt{n!}}\,|n\>\;. \e
Observe that in the present case it suffices that $\pi\le\gamma<\pi$ to 
achieve the needed orthogonality. Reassuringly, therefore, we have been 
able to deduce from our several postulates that the canonical coherent 
states are the unique family of coherent states associated with the 
harmonic oscillator dynamics.

\section*{Application to Hydrogen-like Spectrum}
Finding coherent states for the bound state portion of the hydrogen atom 
has been a long-standing problem. Surely, various proposals for such 
coherent states have been made (see, e.g., \cite{help}), and just as 
surely they generally differ one from another. One means to gauge such 
proposals is how well they do in the semi-classical regime, namely, what 
is the spread in the energy levels for highly excited systems. Ideally, 
one would prefer that the spread decreases as the excitation level rises 
so that more nearly classical-like behavior is obtained. A measure of the 
spread is provided by the variance, and therefore it is appropriate to 
focus on the variance in the proposed coherent states. While the full 
hydrogen atom has been treated elsewhere, we content ourselves here with 
a simple one-dimensional model which serves to illustrate the principles 
involved in a clearer fashion. 

We now turn our attention to a one-dimensional model problem with the 
hydrogen-like spectrum
  \b  E_n=\omega[1-1/(n+1)^2] \;.  \e
In this case
  \b  \rho_n=\prod_{l=1}^n\s \frac{(l^2+2l)}{(l+1)^2}=\frac{1}{2}
\bigg(\frac{n+2}{n+1}\bigg)\;, \e
and thus the coherent states in question are defined by 
  \b |J,\gamma\>=N(J)^{-1/2}\sum_{n=0}^\infty\sqrt{\bigg(
\frac{2n+2}{n+2}\bigg)}\s J^{n/2}\s e^{-i\gamma[1-(n+1)^{-2}]}\,|n\>\;.  \e
Here
  \b N(J)=\sum_{n=0}^\infty\bigg(\frac{2n+2}{n+2}\bigg)\, J^n = 
\frac{2}{1-J}+\frac{2}{J^2}\s[J+\ln(1-J)]\;,  \e
provided that $0\le J<J^*=1$. As in the general case, these states 
clearly exhibit temporal stability, i.e.,  
 \b  e^{-i\H t}\s|J,\gamma\>=|J,\gamma+\omega t\>\;.  \e  
\subsection*{Variance}
By design, of course, these states fulfill the condition
  \b  \<J,\gamma|\s\H\s|J,\gamma\>=\omega\s J \;. \e
A question of particular interest, however, refers to the {\it variance} 
of the energy in each of the given coherent states since this quantity 
serves to indicate how well the energy is peaked about its mean value 
in the coherent state $|J,\gamma\>$. 

It may be shown (e.g., by direct computation) that for the hydrogen-like 
model under discussion the variance
  \b  &&v(J)\equiv\<J,\gamma|\s\H^2\s|J,\gamma\>-\<J,\gamma|\s\H\s|
J,\gamma\>^2\\
   &&\hskip.93cm \le (3\omega^2/4)\s J\s(1-J)\;.  \e
It is noteworthy that the variance vanishes not only for $J=0$ but for $J=1$
as well.
This fact implies that the state $|J,\gamma\>$ is peaked in its energy 
values about its mean value when $J\approx 0$ and when $J\approx 1$. 

We now proceed to discuss the variance in a more general fashion.

\section*{Variances for more General Systems}
Let us discuss the variance for rather general systems for which $J^*=1$.
This analysis leads to further information about the hydrogen-like model 
as well as many other examples. 

In the general case, the energy variance is defined by
 \b && v(J)=\<J,\gamma|\s\H^2\s|J,\gamma\>-\<J,\gamma|\s\H\s|J,\gamma\>^2\\
&&\hskip.91cm=\frac{\Sigma e_n^2\s J^n/\rho_n}{\Sigma\s J^n/\rho_n}-
\frac{(\Sigma\s e_n\s J^n/\rho_n)^2}{(\Sigma\s J^n/\rho_n)^2} \\
&&\hskip.91cm=\frac{1}{2}\s\frac{\Sigma_{n,m}\s(e_n-e_m)^2\s J^{n+m}/
\rho_n\rho_m}{\Sigma_{n,m}\s J^{n+m}/\rho_n\rho_m}\;. \e
Let us examine $v(J)$ at the two extremes $J\approx0$ and $J\approx1$. 

First, for $J\approx0$, we readily see that
  \b  v(J)=e_1\s J +O(J^2) \;. \e
In short, 
   \b v(J)\propto J  \e
near $J=0$. 
 
For $J\approx1$ the analysis is somewhat more involved. We note that 
$v(J)$ may be written as 
 \b  v(J)=\frac{1}{2}\s\frac{\Sigma\s J^m/\rho_m\,\Sigma\s(\delta_n-
\delta_m)^2\s J^n/\rho_n}{\Sigma\s J^m\rho_m\,\Sigma\s J^n/\rho_n}\;, \e
where $\delta_n\equiv1-e_n$. Observe that $\delta_n\ra0$ as $n\ra\infty$. 
For the moment we assume even more, namely, that 
$\Sigma\s\delta_m^2<\infty$. Since large $n$ values dominate the 
$n$-sums in the numerator and the denominator, then near $J=1$ it 
suffices to consider
  \b &&v(J) =\frac{1}{2}\s\frac{\Sigma\s\delta_m^2\s J^m/\rho_m\,
\Sigma\s J^n/\rho_\infty}{\Sigma\s J^m/\rho_\infty\,\Sigma\s J^n/
\rho_\infty} \\
  &&\hskip.85cm=(1-J)\s(\half\s\rho_\infty\s\Sigma\s\delta_m^2/\rho_m)
+O([1-J]^2)\;.  \e
Roughly speaking, if $\delta_m^2\propto m^{-\tau}$, for large $m$, 
$1<\tau$, then we have shown to leading order that
    \b v(J)\propto (1-J)  \e
near $J=1$.
On the other hand, if $\delta_m^2\propto m^{-\tau}$, for large $m$, 
$0<\tau<1$, it follows to leading order that
  \b  v(J)\propto (1-J)^\tau   \e
near $J=1$.

Finally,  we learn that the vanishing of the variance for large quantum 
numbers, i.e., when $J\approx 1$ is a rather general phenomena, given 
the choice of coherent states to which we have been led in the present 
article. This fact would seem to confirm their utility in semi-classical 
analyses rather generally.

\section*{Related work}
Several other papers have recently appeared dealing with topics raised in 
this article, and the interested reader may wish to consult them directly. 
In \cite{ant} temporally stable coherent states are developed for the 
infinite square well and for the P\"oschl-Teller potential. A review 
of various attempts to develop coherent states in general and hydrogen 
atom coherent states in particular is given in \cite{cra}. 

\section*{Acknowledgements}
I take this opportunity to thank several colleagues who have been 
recently involved with the author in one way or another regarding the 
subject of coherent states. These collaborators are: J.-P. Antoine, 
B. Bodmann, J.-P. Gazeau, P. Monceau, K.A. Penson, J.-M. Sixdeniers, 
S.V. Shabanov, and G. Watson.


\begin{thebibliography}{99}
\bibitem{kla2} J.R. Klauder, ``The Action Option and the Feynman 
Quantization of Spinor Fields in 
Terms of Ordinary C-Numbers", Annals of Physics {\bf 11}, 123-168 (1960).
\bibitem{barg} V. Bargmann, ``On a Hilbert Space of Analytic Functions 
and an
Associated Integral Transform, Part I'', Commun. Pure and Applied Math. 
{\bf 14}, 187-214 (1961).
\bibitem{gla} R.J. Glauber, ``The Quantum Theory of Optical Coherence'', 
Phys. Rev. {\bf 130}, 2529-2539 (1963).
\bibitem{kla19}J.R. Klauder, ``Continuous-Representation Theory II.  
Generalized Relation Between 
Quantum and Classical Dynamics", J. Math. Phys. {\bf 4}, 1058-1073 (1963).
\bibitem{per}A.M. Perelomov, ``Coherent States for Arbitrary Lie Groups'', 
Commun. Math. Phys. {\bf 26}, 222-236 (1972).
\bibitem{gil}R. Gilmore, ``On the Properties of Coherent States'', 
Revista Mexicana de Fisica {\bf 23}, 143-187 (1974).
\bibitem{bar} A.O. Barut and L. Girardello, ``New `Coherent States' 
Associated with Non-Compact Groups'', Commun. Math.Phys. {\bf 21}, 
41-55 (1972).
\bibitem{nie} M.M. Nieto and L.M. Simmons, Jr., ``Coherent States for 
General Potentials'', Phys. Rev. Lett. {\bf 41}, 207-210 (1978). 
\bibitem{kla18}J.R. Klauder, ``Continuous-Representation Theory I.  
Postulates of 
Continuous Representation Theory", J. Math. Phys. {\bf 4}, 1055-1058 (1963).
\bibitem{kla222}J.R. Klauder, ``Coherent State Path Integrals {\it without} 
Resolutions of Unity'', Found. Phys. {\bf 31} 57-67 (2001).
\bibitem{kla191}J.R. Klauder, ``Coherent States for the Hydrogen Atom.'' 
J. Phys. A: Math. Gen. {\bf 29}, L293-L298 (1996).
\bibitem{kla37}J.R. Klauder, ``Continuous-Representation Theory III.  On 
Functional Quantization of Classical Systems", 
J. Math. Phys. {\bf 5}, 177-187 (1964).
\bibitem{kla207}J.-P. Gazeau and J.R. Klauder, ``Coherent States for 
Systems with Discrete and Continuous Spectrum'', J. Phys. A: Math. 
Gen. {\bf  32}, 123-132, (1999).
\bibitem{help}L.S. Brown, Am. J. Phys. {\bf 41}, 525 (1973); J. Mostowski, 
Lett. Math. Phys. {\bf 2}, 1 (1977); J.C. Gay, D. Delande, and A. Bommier, 
Phys. Rev. A {\bf 39}, 6587 (1989); M. Nauenberg, Phys. Rev. A {\bf 40}, 
1133 (1989); Z.D. Gaeta and C.R. Stroud, Jr., Phys. Rev. A {\bf 42}, 6308 
(1990); J.A. Yeazell and C.R. Stroud, Jr., Phys. Rev. A {\bf 43}, 5153 
(1991); M. Nauenberg, in {\it Coherent States: Past, Present, and Future}, 
Eds. D.H. Feng, J.R. Klauder, and M.R. Strayer (World Scientific, 
Singapore, 1994), p.345; I. Zlatev, W.-M. Zhang, and D.H. Feng, Phys. 
Rev. {\bf 50}, R1973 (1994); R. Bluhm, V.A. Kostelcky, and B. Tudose, 
(1995); J.R. Klauder, J. Phys. A: Math. Gen. {\bf 29}, L293-L298 (1996); 
P. Majumdar and H.S. Sharatchandra, Phys. Rev. A {\bf 56}, R3322 (1997); 
M.G.A. Crawford,  Phys. Rev. A {\bf 62}, 012104, 1-7 (2000).
\bibitem{ant}J.-P. Antoine, J.-P. Gazeau, P. Monceau, J.R. Klauder,  and 
K.A. Penson, ``Temporally Stable Coherent States for Infinite Well 
and P\"oschl-Teller Potentials'', J. Math. Phys. {\bf 42}, 2349-2386 
(2001).
\bibitem{cra}M.G.A. Crawford, ``Temporally Stable Coherent States in 
Energy-degenerate Systems: The Hydrogen Atom'', 
Phys. Rev. A {\bf 62}, 012104, 1-7 (2000). 
%\bibitem{kla}
%\bibitem{kla}



\end{thebibliography}
\end{document}